\documentstyle[preprint,revtex]{aps}
\tightenlines
\begin{document}
\draft
\begin{title}
Quartets of superdeformed bands and supersymmetry breaking
\end{title}
\author{
R.D.~Amado$^{(1),(2)}$, R.~Bijker$^{(3)}$, F.~Cannata$^{(1),(4)}$, %
J.P.~Dedonder$^{(1),(5)}$ and N.R.~Walet$^{(2)}$ }
\begin{instit}
$^{(1)}$ Paul Scherrer Institute, CH-5232 Villigen-PSI, Switzerland \\
$^{(2)}$ Department of Physics, University of Pennsylvania,
Philadelphia, PA 19104, USA$^{(a)}$ \\
$^{(3)}$ R.J. Van de Graaff Laboratory, University of Utrecht,
P.O. Box 80000,
NL-3508 TA Utrecht, The Netherlands \\
$^{(4)}$ Dipartimento di Fisica and INFN,
I-40126 Bologna, Italy$^{(b)}$ \\
$^{(5)}$ Laboratoire de Physique
Nucl\'eaire, Universit\'e Paris 7,
2 Place Jussieu, F-75251 Paris Cedex 05 and
Division de  Physique Th\'eorique,
IPN, F-91406 Orsay, France$^{(c)}$
\end{instit}

\begin{abstract}
We examine the prediction of supersymmetric quantum mechanics that
bands with identical gamma-ray energies occur in quartets.
The experimental data suggest that this scenario is actually
realized in nature. In the $A=150$ mass region, four known pairs
of isospectral bands can be grouped in two quartets, while there are
indications of such patterns around $A=190$. We
introduce a small supersymmetry breaking,
necessary to describe the details of the data. We derive relations
among the transition rates that can be used to test our predictions.
\end{abstract}

\section{Introduction}

In a recent letter \cite{ABCD} we showed that the remarkable
identical gamma ray bands (isospectral gamma rays)
seen in neighboring superdeformed nuclei \cite{Byrski,Twin} have
a natural description in terms of supersymmetric quantum
mechanics (SSQM).
The role of supersymmetry in superdeformed nuclei has not in general
been emphasized except in \cite{Gelberg}.
The SSQM picture requires that bands with isospectral gamma
rays in an even and neighboring odd nucleus be accompanied by
two more isospectral bands, one in the odd nucleus
and another in a neighboring even nucleus. As we have shown
in \cite{ABCD} this is most naturally realized when the even nuclei
differ by two nucleons.
We also suggested that two of the bands actually occurred at a very high
excitation energy, so that they would be unobservable in experiment.
In this paper we point out that all four band might actually be
observable.
We show that the experimentally observed isospectral pairs can very
naturally be grouped together in two quartets.
We point out that two such
quartets of gamma ray sequences have indeed been observed in
superdeformed nuclei in the $A=150$ mass region, although to our
knowledge
they have seldom been associated together \cite{Zuber,Rag}.
 As could be expected,
the quartets are not
perfectly isospectral which suggests a small breaking
of the supersymmetry.

In this paper we devise a simple form of supersymmetry breaking and
show that this accounts for the bulk of the experimental data related
to these quartets.
Our work on symmetry breaking gives promise of making a
connection between our approach and more microscopic ones.
Because SSQM relates the four gamma ray cascades that make up
the quartet, it also makes predictions about transition rates
in the four bands. These are presented here to stimulate experimental
interest when the new facilities to study superdeformed nuclei come
on line \cite{Beck}.

\section{Gamma ray quartets in superdeformed nuclei}

The recently observed sequences of isospectral gamma rays in
neighboring even and odd nuclei lead naturally to an interpretation
in terms of
supersymmetric quantum mechanics. As is well known SSQM is designed to
have equal energy eigenvalues for bosonic (even nucleus) and fermionic
(odd nucleus) states. In addition SSQM requires that the dimension of
the bosonic part
and the fermionic part of a given degenerate multiplet be identical.
For the case of superdeformed nuclei this implies that for a given
excitation energy the number of states in the even nucleus equals that
in the neighboring odd nucleus.
If the state in the even nucleus has angular momentum $L$ (integer) and
the corresponding state in the odd nucleus $J$ (half integer), the
degeneracies are $2L+1$ and $2J+1$, respectively. Of course these two
numbers can never be equal. In \cite{ABCD} we showed that the state
counting problem can be solved in the following way.
Suppose the state in the odd nucleus is obtained by
coupling one particle (or hole) with spin or pseudospin \cite{pseu}
$s=\frac{1}{2}$ to the corresponding state in the even
nucleus. The angular momenta in the odd system are then
$J_>=L+\frac{1}{2}$ and $J_<=L-\frac{1}{2}$.
Suppose further that in the even nucleus with two particles
(or two holes) more than in the original even nucleus
there is a level with angular momentum $L$.
Now there are $2(2L+1)$ even states and precisely the same number of
odd states. These four states form a degenerate SSQM quartet.
Transitions between sets of these states will lead to four isospectral
gamma ray sequences, one in the even nucleus $A$, one
in the other even nucleus $A \pm 2$, and two in the
odd $ A \pm 1$ nucleus. One of these last two is among the
$J_>$ states and the other among the $J_<$ states. As we
shall show below cross-over transitions between
$J_<$ and $J_>$ are very small.

It is this quartet of gamma rays sequences that
is the hallmark of SSQM. Let us examine two candidates.
In Tables I and II we show that all four known isospectral
pairs can be grouped in two of these quartets.
In Table I we compare gamma ray sequences in
four superdeformed bands in the adjacent nuclei
$^{152}$Dy, $^{151}$Tb (twice) and $^{150}$Gd
\cite{Byrski,Twin86,Bentley87,Fallon89,sd}.
As is well known spins have not been measured yet, and we have
thus chosen spin
assignments consistent with a SSQM interpretation. We have followed the
spin assignments in the compilation \cite{sd} as closely as possible.
We see that the four bands are very
close in energy, but are not completely degenerate.
In spite of the breaking of supersymmetry it is clear that the
quartet bears the strong family resemblance suggested
by SSQM. A second example of such a quartet
is shown in Table II .  Here the four
bands are in $^{148}$Gd, $^{147}$Gd (twice), and
$^{146}$Gd \cite{Zuber,Rag,sd,Hebbinghaus90,Deleplanque88}.
Again the spins have been assigned to stress
the SSQM connection, and again we see the general pattern
of slightly broken SSQM. The pattern does seem
to shift around $L=46$, which is due to a level crossing
that is not contained in the simple
SSQM quartet picture. The crossing occurs in two bands
that do not form an isospectral pair:
the 1b-band in $^{146}$Gd and the 1b-band in
$^{147}$Gd. Before the level crossing the last band is isospectral
with the 1b-band in $^{148}$Gd, and
the first with the 2b-band in $^{147}$Gd.
As can be seen from Fig.~2 the nature of the crossing is very similar
in both cases. Thus the band crossing stresses the correlation between
 the two isospectral pairs within the quartet.
It suggests that a description in terms of
two independent pairs of superdeformed bands is not adequate.

It would certainly be interesting to pursue the search for identical
superdeformed bands \cite{Casten} and investigate whether more
supersymmetric quartets exist. There are preliminary indications
that such quartets are present in the $A=190$ mass region
(see Table III). Furthermore, it is very important to determine
the spins experimentally. The spin assignments of SSQM are a
central feature of the theory and their verification
or refutation is crucial to test the application of
supersymmetry to superdeformed nuclei. Note that
the SSQM spin assignments do not always agree with other
phenomenological determinations\cite{Zeng9192,Draper90}.

\section{SSQM breaking}

In a SSQM description the quartet of states discussed in Section II
is degenerate. This degeneracy implies that the energy only depends
on $L$. A small spin-orbit splitting will break the supersymmetry.
At the algebraic level of our treatment it is irrelevant whether
this is a spin-orbit or a pseudo-spin-orbit coupling.
For each nucleus, we take a Hamiltonian of the form
\begin{equation}
H_i = a_i \; {\bf L} \cdot {\bf L} + b \; {\bf s}\cdot {\bf L} ~,
\end{equation}
where ${\bf L}$ is the angular momentum of the core and ${\bf s}$
is the spin (or pseudo-spin) of the odd particle (hole).
The coefficients $a_i$ are inversely proportional to the moment
of inertia in each nucleus. The strength of the spin-orbit coupling,
$b$, is taken the same for each nucleus. In the sequences
we are considering (Tables I and II) we label the three nuclei by
$i = 0$ for $^{152}$Dy or $^{148}$Gd, by $i = 1$ for $^{151}$Tb
or $^{147}$Gd and by $i = 2$ for $^{150}$Gd or $^{146}$Gd.

Supersymmetry implies that all the $a_i$'s are equal. A plot of
the gamma ray energies (see Figures I and II) against the core
angular-momentum $L$ confirms this
by revealing a near perfect straight line with almost the same slope
for all transitions. With $a_0$ as the reference, we assume small $L$
dependent departures in $a_1$ which we parametrize as
$a_1 = a_0 + \frac{\epsilon}{L}$ \cite{VMI}.
If this departure is due to the extra hole (or extra particle),
we expect $a_2 = a_0 + \frac{2 \epsilon}{L}$. With these expressions
we find for the transition energies $E_{\gamma i}(I)=E_i(I)-E_i(I-2)$,
where $E_i(I)$ is the excitation energy of the state with
angular momentum
$I$,
\begin{eqnarray}
E_{\gamma 0}(L) &=& a_0 (4 L -2) ~,
\nonumber\\
E_{\gamma 1}(L \pm {\textstyle \frac{1}{2}}) &=&
a_0 (4 L -2) + 2 \epsilon \pm b ~,
\nonumber\\
E_{\gamma 2}(L) &=& a_0 (4 L -2) + 4 \epsilon ~.
\end{eqnarray}
In this parametrization
\begin{eqnarray}
\Delta E_{\gamma 01} &=&
E_{\gamma 0}(L)-E_{\gamma 1}(L+{\textstyle \frac{1}{2}})=
\Delta E_{\gamma 12}
\nonumber\\
&=& E_{\gamma 1}(L-{\textstyle \frac{1}{2}})-E_{\gamma 2}(L)
= 2 \epsilon + b ~.
\end{eqnarray}
Both in the case of Table I and
Table II this difference is essentially zero.
This relationship among the gamma ray sequences in the
Gadolinium isotopes has been noted by Ragnarsson \cite{Rag}
but not in the context of SSQM and with different spin assignments.
(We stop the comparison  at $L=46$ in Table II
as discussed in the previous Section).
We have no simple explanation for why
$E_{\gamma 1}(L+\frac{1}{2})=E_{\gamma 0}(L)$, but we see that for
both Tables our parametrization immediately gives that if
$E_{\gamma 1}(L+\frac{1}{2})=E_{\gamma 0}(L)$,
then $E_{\gamma 2}(L)=E_{\gamma 1}(L-\frac{1}{2})$ as seen.
The parametrization also implies that
$E_{\gamma 2}(L)-E_{\gamma 0}(L)= 4 \epsilon$ independent of $L$
and that too is fairly well born out in the Tables.
We find for the nuclei of Table I, $\epsilon = -b/2=-6.3 \pm 1.3$ keV,
which should be compared to $a_0 = 47 \pm 1$ keV. Thus the term
proportional
to $\epsilon$ is indeed a small perturbation in the expression for
$a_2$.
Similarly, for  Table II we find $\epsilon = -b/2 = 8.3 \pm 0.6$ keV
and $a_0= 55 \pm 3$.

We see that a very simple approach to SSQM breaking based on a weak
spin-orbit coupling and a slight difference in how moments of
inertia change with added particles or holes accounts for the data of
Tables I and II. Although at the moment we do not have a microscopic
understanding of the pair wise equality of gamma ray energies,
$E_{\gamma 1}(L+\frac{1}{2})=E_{\gamma 0}(L)$ and
$E_{\gamma 1}(L-\frac{1}{2})=E_{\gamma 2}(L)$, we have shown
a mechanism in terms of a broken supersymmetry that exhibits this
pattern.
Whether the remaining degeneracy is accidental or the result
of a deeper symmetry remains to be seen.
It is our goal in this paper to show that we can realize the pattern
of  super-symmetry breaking apparent in the experimental data.
\section{Transition strengths}

In the tentative spin assignments in Table I and II we have assumed
that the subsequent states in a superdeformed band differ by two units
of angular momentum.
If we further adopt a pseudo-spin picture
in which the $E2$ transition operator is independent of the
pseudo-spin, the ratio between the four possible intraband
transition probabilities only depends on geometric factors from
angular momentum algebra \cite{BK}.
Again taking the $B(E2)$ value in the $A$ nucleus as a reference,
we find the following relations for the intraband transitions
\begin{eqnarray}
B(E2;J_>=L+{\textstyle \frac{5}{2}} \rightarrow J_>=
L+{\textstyle \frac{1}{2}}) &=&
B(E2;L+2 \rightarrow L) ~,
\nonumber\\
B(E2;J_<=L+{\textstyle \frac{3}{2}} \rightarrow J_<=
L-{\textstyle \frac{1}{2}}) &=&
\frac{L(2L+5)}{(2L+1)(L+2)}
\nonumber\\
&& \times B(E2;L+2 \rightarrow L) ~.
\end{eqnarray}
The transition probabilities in the even nuclei in the quartet
are identical in this scheme. For typical values of the angular momenta
in superdeformed bands (see Tables I and II) the intraband
transitions energies are nearly the same
(isospectral). The related interband transition
between the two bands in the odd nucleus,
\begin{eqnarray}
B(E2;J_<=L+{\textstyle \frac{3}{2}} \rightarrow J_>=
L+{\textstyle \frac{1}{2}}) &=&
\frac{2}{(2L+1)(L+2)}
\nonumber\\
&& \times B(E2;L+2 \rightarrow L) ~,
\end{eqnarray}
is highly suppressed with respect to the intraband transitions
for typical values of the angular momentum. The other
interband transition from $J_> =L+\frac{5}{2}$ to $J_< =L-\frac{1}{2}$
is strictly forbidden for quadrupole radiation.

Following the notation of the previous section we label
the bands by $i=0,1\pm,2$. The ratio of the transition probabilities
$B_{ij}$ only depends on a simple geometric factor
$B_{00} : B_{22} : B_{1+1+} : B_{1-1-} : B_{1-1+} = 1 : 1 : 1 :
\frac{L(2L+5)}{(2L+1)(L+2)} : \frac{2}{(2L+1)(L+2)}$.
For typical values of $L$ in superdeformed bands it is appropriate
to take the large $L$ limit for these ratios, giving
$1 : 1 : 1 : 1 : \frac{1}{L^2}$, which shows that for large $L$
all intraband transition probabilities are equal and that the
interband transition from $J_<=L+\frac{3}{2}$ to $J_>=L+\frac{1}{2}$
is down by $\frac{1}{L^2}$, a very big suppression.

The other interband transition from $J_>=L+\frac{1}{2}$ to
$J_<=L-\frac{1}{2}$ depends on the same matrix element as the
static quadrupole moments.
With the quadrupole moment $Q_0(L)$ in the $A$ nucleus as a reference
we find
\begin{eqnarray}
Q_1(L+{\textstyle \frac{1}{2}}) &=& Q_0(L) ~,
\nonumber\\
Q_1(L-{\textstyle \frac{1}{2}}) &=&
\frac{(L-1)(2L+3)}{L(2L+1)} \; Q_0(L) ~.
\end{eqnarray}
The ratio of the quadrupole moments of the states in the four
superdeformed bands is then
$Q_0(L) : Q_2(L) : Q_1(L+\frac{1}{2}) : Q_1(L-\frac{1}{2}) =
1 : 1 : 1 : \frac{(L-1)(2L+3)}{L(2L+1)}$, which shows that in the
large $L$ limit the quadrupole moments of the members of a quartet are
identical.
The $E2$ transition between the two states in the odd nucleus
belonging to the same quartet is suppressed by $1/L^2$,
\begin{eqnarray}
B(E2;J_>=L+{\textstyle \frac{1}{2}} \rightarrow J_<=
L-{\textstyle \frac{1}{2}}) =
\frac{3}{(2L+1)(L+1)} \; B(E2;L \rightarrow L) ~.
\end{eqnarray}
The corresponding gamma ray energy for this transition is determined by
the spin-orbit splitting, $E_{\gamma}=\frac{1}{2} b(2L+1)$.
The present data on lifetimes and quadrupole moments for superdeformed
bands is still quite scarce
\cite{Hebbinghaus90,Deleplanque88,Moore90,Willsau92,Bentley87} and
it will be certainly interesting to check the above predictions when
the technology develops to the point of accurate measurements of
lifetimes.

We stress that our predictions do not depend on the details of
the dynamics, but only on the SSQM prediction that the four states have
nearly identical internal structures and therefore have the same
reduced
matrix elements for quadrupole transitions. They further depend on the
assumption
that the odd states are constructed by coupling a spin (or pseudo-spin)
$s=\frac{1}{2}$ to the orbital state.
The symmetry breaking considered in the previous section is diagonal
and does not affect the transition strengths.
In spite of the weak nature of all
these assumptions the connection of quartets across even and odd nuclei
makes interesting and verifiable predictions.

These prediction should hold to high accuracy, and should not be
confused
with the much weaker relations that can be derived in the
Bohr-Mottelson model
for nuclei with roughly equal deformation.

\section{Prospects and summary}

Supersymmetric quantum mechanics
relates states in odd and even neighboring nuclei. It
describes a quartet of states across nuclei with $A$ (even)
nucleons, $(A + 1)$ and $(A + 2)$ nucleons that are degenerate and
hence lead to four isospectral bands in the nuclei, one each in the
$A$ and $(A + 2)$ nuclei and two in the $(A + 1)$ nucleus.
Although, we pointed out \cite{ABCD} that other realizations
of SSQM are possible, it seems that the $A$, $(A + 1)$, $(A + 2)$
scheme is not only the most natural, but is actually realized in
nature.
 In this paper we have shown that two examples
of such quartets appear to exist in the data although they have
seldom been regrouped to emphasize the SSQM connection.
It may very well be that these quartets of gamma ray sequences
in superdeformed nuclei are the best example of supersymmetry
so far discovered in physics.

We have tried to realize the breaking of supersymmetry in the simplest
possible way, in order to describe the experimental data. We have not
attempted
to derive the symmetry from an underlying microscopic Hamiltonian. Such
an approach would be complementary to our approach, but it seems highly
unlikely that current mean-field techniques can give the required
accuracy.

Our approach makes definite assumptions about the spins of the related
states and predicts equal transition strengths across the
quartets. New experiments with EUROBALL and GAMMASPHERE could be very
helpful in checking the spin assignments and the predictions for the
transition strengths, as well as in discovering possible new examples
of quartets of superdeformed bands. As we have already stated, there
are indications of such quartets in the $A=190$ mass region. One can
look
either for isotopic or isotonic quartets. The presently available
data \cite{sd} suggests two examples. The better of the two is an
isotopic $Z=80$ quartet
composed of the superdeformed band in $^{192}$Hg \cite{Ye}, two of
the four known bands ($^{193}$Hg(2) and $^{193}$Hg(3), \cite{Cullen})
in $^{193}$Hg and one of the three known bands ($^{194}$Hg(3),
\cite{Riley90,Beausang90,Stephens90}) in $^{194}$Hg. This is
displayed in Table III where the equality between the quantities
$\Delta E_{\gamma 01}$ and $\Delta E_{\gamma 12}$ holds within the
experimental
uncertainties. We obtain $\epsilon = -0.05 \pm 0.48 $ keV and
$b = 8.85\pm 2.35$ keV. Note also that by shifting the spin of the last
two bands up by two units, the pattern remains the same but the values
of the symmetry breaking parameters $\epsilon$ and $b$ are changed
to $9.3 \pm 0.42$ keV and $-9.85 \pm 2.19$ keV,
respectively.
An other possibility may be the isotonic $N = 112$
quartet made of $^{192}$Hg \cite{Ye}, $^{193}$Tl(a), $^{193}$Tl(b)
\cite{fernand} and $^{194}$Pb \cite{brinkman}
though, there, the difference between $\Delta
E_{\gamma 01}$ and $\Delta E_{\gamma 12}$ is beyond experimental errors
and cannot be accommodated by the simple SSQM breaking
scheme of Section III.
However the measured quadrupole moments
\cite{Moore90,Willsau92,Bentley87} for the superdeformed
bands in $^{192}$Hg and $^{194}$Pb agree within experimental errors.
The balance of boson and fermion degrees of freedom, which
was discussed in Section II and in \cite{ABCD}, implies that these two
examples which
involve the same band in the $^{192}$Hg nucleus should be incompatible
unless there are two (almost) degenerate bands in $^{192}$Hg or unlesss
the supersymmetry multiplet is otherwise  enlarged.  At the
present time only one superdeformed band has been observed in
$^{192}$Hg
\cite{Ye}.
Further investigations around mass $192$ would be very valuable
to clarify these points.

To summarize, our SSQM description is not cast in terms of microscopic
variables. Some of the dynamics comes in via the symmetry breaking.
The quartet multiplets are not perfectly isospectral. Most
of this symmetry breaking can be accounted for by a small (pseudo-)spin
orbit term and by some very simple $L$-dependent change in the moments
of inertia. These correction terms should begin to give some insight
into the connection between SSQM dynamics and the more conventional
dynamical avenues \cite{bunch,Stephens2,Stephens90}.

\acknowledgments

RDA, FC, and JPD  would like to thank the Paul Scherrer Institute
and Milan Locher for
bringing us together and for providing a very pleasant environment
in which much of this work was done.  RDA and NRW are
supported in part by the U.S. National Science Foundation and
RB by the Stichting voor Fundamenteel Onderzoek der Materie (FOM)
with financial support from the Nederlandse Organisatie voor
Wetenschappelijk Onderzoek (NWO).
The Division de Physique Th\'eorique is a Research Unit of the
Universities Paris 11 and 6 associated to CNRS.

\newpage
\figure
{Gamma ray energies versus the core angular momentum $L$
for the quartet of superdeformed bands in $^{152}$Dy (solid),
$^{151}$Tb (2b) (dashed) and (1b) (dotted), and
$^{150}$Gd (2b) (dashed-dotted). On the scale of the figure the first
two and the last two are indistinguishable.}
\figure
{Gamma ray energies versus the core angular momentum $L$
for the quartet of superdeformed bands in $^{148}$Gd (1b) (solid),
$^{147}$Gd (1b) (dashed) and (2b) (dotted), and
$^{146}$Gd (1b) (dashed-dotted).}
\newpage
\begin{table}
\caption{Quartet of superdeformed bands in $^{150}$Gd, $^{151}$Tb
and $^{152}$Dy.
Spin assignments are from \protect{\cite{sd}}, except
for a shift of two units
of angular momentum in the bands $^{151}$Tb(1b) and $^{150}$Gd(2b).
All energies are in keV.}
\begin{tabular}{crcrrcrcrrr}
\multicolumn{2}{c}{$^{152}$Dy} &
\multicolumn{2}{c}{$^{151}$Tb(2b)} & &
\multicolumn{2}{c}{$^{151}$Tb(1b)} &
\multicolumn{2}{c}{$^{150}$Gd(2b)} & & \\
\tableline
$L$ & $E_{\gamma 0}$ &
$J$ & $E_{\gamma 1}$ & $\Delta E_{\gamma 01}$ &
$J$ & $E_{\gamma 1}$ &
$L$ & $E_{\gamma 2}$ & $\Delta E_{\gamma 12}$ &
$\Delta E_{\gamma 20}$ \\
\tableline
24 & 602.3  &      &      &      &      &        &    &      & & \\
26 & 647.2  & 26.5 & 647  &  0.2 &      &        &    &      & & \\
28 & 692.2  & 28.5 & 692  &  0.2 &      &        &    &      & & \\
30 & 737.5  & 30.5 & 738  & -0.5 & 29.5 & 728.0  &    &      & & \\
32 & 783.5  & 32.5 & 783  &  0.5 & 31.5 & 769.2  & 32 & 770  & -0.8
& -13.5 \\
34 & 829.2  & 34.5 & 828  &  1.2 & 33.5 & 811.3  & 34 & 813  & -1.7
& -16.2 \\
36 & 876.1  & 36.5 & 876  &  0.1 & 35.5 & 854.0  & 36 & 856  & -2.0
& -20.1 \\
38 & 923.1  & 38.5 & 922  &  1.1 & 37.5 & 898.0  & 38 & 900  & -2.0
& -23.1 \\
40 & 970.0  & 40.5 & 970  &  0.0 & 39.5 & 942.8  & 40 & 944  & -1.2
& -26.0 \\
42 & 1017.0 & 42.5 & 1016 &  1.0 & 41.5 & 988.7  & 42 & 990  & -1.3
& -27.0 \\
44 & 1064.8 & 44.5 & 1063 &  1.8 & 43.5 & 1034.9 & 44 & 1036 & -1.1
& -28.8 \\
46 & 1112.7 & 46.5 & 1112 &  0.7 & 45.5 & 1082.5 & 46 & 1082 &  0.5
& -30.7 \\
48 & 1160.8 & 48.5 & 1158 &  2.8 & 47.5 & 1130.2 & 48 & 1130 &  0.2
& -30.8 \\
50 & 1208.7 & 50.5 & 1207 &  1.7 & 49.5 & 1178.9 & 50 & 1180 & -1.1
& -28.7 \\
52 & 1256.6 & 52.5 & 1256 &  0.6 & 51.5 & 1228.5 & 52 & 1229 & -0.5
& -27.6 \\
54 & 1304.7 & 54.5 & 1305 & -0.3 & 53.5 & 1278.5 & 54 & 1277 &  1.5
& -27.7 \\
56 & 1353.0 & 56.5 & 1353 &  0.0 & 55.5 & 1330.0 & 56 & 1327 &  3.0
& -26.0 \\
58 & 1401.7 &      &      &      & 57.5 & 1380.7 & 58 & 1378 &  2.7
& -23.7 \\
60 & 1449.4 &      &      &      & 59.5 & 1432.5 &    &      & & \\
\end{tabular}
\end{table}

\newpage
\begin{table}
\caption{Quartet of superdeformed bands in $^{146-148}$Gd.
Spin assignments are from \protect{\cite{sd}},
except for a shift of two units
of angular momentum in the bands $^{147}$Gd(2b) and $^{146}$Gd(1b).
All energies are in keV.}
\begin{tabular}{crcrrcrcrrr}
\multicolumn{2}{c}{$^{148}$Gd(1b)} &
\multicolumn{2}{c}{$^{147}$Gd(1b)} & &
\multicolumn{2}{c}{$^{147}$Gd(2b)} &
\multicolumn{2}{c}{$^{146}$Gd(1b)} & & \\
\tableline
$L$ & $E_{\gamma 0}$ &
$J$ & $E_{\gamma 1}$ & $\Delta E_{\gamma 01}$ &
$J$ & $E_{\gamma 1}$ &
$L$ & $E_{\gamma 2}$ & $\Delta E_{\gamma 12}$ &
$\Delta E_{\gamma 20}$ \\
\tableline
26 &  700 & 26.5 &  696.3 &  3.7 & 25.5 &  731.2 &    &        & & \\
28 &  747 & 28.5 &  745.7 &  1.3 & 27.5 &  779.1 &    &        & & \\
30 &  797 & 30.5 &  795.4 &  1.6 & 29.5 &  826.9 & 30 &  826.7 &  0.2
& 29.7 \\
32 &  847 & 32.5 &  846.7 &  0.3 & 31.5 &  877.2 & 32 &  878.5 & -1.3
& 29.5 \\
34 &  898 & 34.5 &  899.9 & -1.9 & 33.5 &  928.6 & 34 &  929.4 & -0.8
& 31.4 \\
36 &  950 & 36.5 &  954.4 & -4.4 & 35.5 &  981.5 & 36 &  983.1 & -1.6
& 33.1 \\
38 & 1004 & 38.5 & 1009.2 & -5.2 & 37.5 & 1035.3 & 38 & 1038.8 & -3.5
& 34.8 \\
40 & 1058 & 40.5 & 1065.3 & -7.3 & 39.5 & 1090.3 & 40 & 1093.9 & -3.6
& 35.9 \\
42 & 1114 & 42.5 & 1120.4 & -6.4 & 41.5 & 1146.4 & 42 & 1148.7 & -2.3
& 34.7 \\
44 & 1171 & 44.5 & 1175.2 & -4.2 & 43.5 & 1203.1 & 44 & 1201.4 &  1.7
& 30.4 \\
46 & 1228 & 46.5 & 1228.6 & -0.6 & 45.5 & 1261.0 & 46 & 1250.3 & 10.7
& 22.3 \\
48 & 1285 & 48.5 & 1277.0 &  8.0 & 47.5 & 1319.0 & 48 & 1297.4 & 21.6
& 12.4 \\
50 & 1344 & 50.5 & 1323.3 & 20.7 & 49.5 & 1378.6 & 50 & 1345.7 & 32.9
& 1.7 \\
52 & 1403 & 52.5 & 1367.1 & 35.9 & 51.5 & 1438.8 & 52 & 1397.0 & 41.8
& -6.0 \\
54 & 1462 & 54.5 & 1413.7 & 48.3 & 53.5 & 1500.0 & 54 & 1448.3 & 51.7
& -13.7 \\
56 & 1520 & 56.5 & 1463.0 & 57.0 & 55.5 & 1559.0 &    &        & & \\
58 & 1580 & 58.5 & 1516.0 & 64.0 &      &        &    &        & & \\
\end{tabular}
\end{table}

\newpage

\begin{table}
\caption{Suggested quartet classification of superdeformed bands
in the isotopes $Z = 80$ for $ A = 192$ to $194$.
Spin assignments are made to fulfill SSQM constraints.
All energies are in keV.}
\begin{tabular}{crcrrcrcrrr}
\multicolumn{2}{c}{$^{192}$Hg} &
\multicolumn{2}{c}{$^{193}$Hg(2)} & &
\multicolumn{2}{c}{$^{193}$Hg(3)} &
\multicolumn{2}{c}{$^{194}$Hg(3)} & & \\
\tableline
$L$ & $E_{\gamma 0}$ &
$J$ & $E_{\gamma 1}$ & $\Delta E_{\gamma 01}$ &
$J$ & $E_{\gamma 1}$ &
$L$ & $E_{\gamma 2}$ & $\Delta E_{\gamma 12}$ &
$\Delta E_{\gamma 20}$ \\
\tableline
10 & 214.6 &      &        &      &  9.5 & 233.7 &  &  & & \\
12 & 257.7 & 12.5 &  254.3 &  3.4 & 11.5 & 274.6 & 12 & 262.3 &  12.3
& -4.6\\
14 & 299.9 & 14.5 &  294.9 &  5.0 & 13.5 & 314.7 & 14 & 302.5 &  12.2
& -2.6\\
16 & 341.1 & 16.5 &  334.2 &  6.9 & 15.5 & 353.5 & 16 & 342.8 &  10.7
& -1.7\\
18 & 381.6 & 18.5 &  374.2 &  7.4 & 17.5 & 392.7 & 18 & 382.1 &  10.6
& -0.5\\
20 & 420.8 & 20.5 &  412.9 &  7.9 & 19.5 & 431.5 & 20 & 420.4 &  11.1
&  0.4\\
22 & 459.1 & 22.5 &  451.0 &  8.1 & 21.5 & 468.7 & 22 & 458.3 &  10.4
&  0.8\\
24 & 496.3 & 24.5 &  488.1 &  8.2 & 23.5 & 504.7 & 24 & 494.6 &  10.1
&  1.7\\
26 & 532.4 & 26.5 &  524.9 &  7.5 & 25.5 & 540.7 & 26 & 531.6 &  9.1
&  0.8\\
28 & 567.9 & 28.5 &  559.9 &  8.0 & 27.5 & 575.4 & 28 & 566.4 &  9.0
&  1.5\\
30 & 602.3 & 30.5 &  595.0 &  7.3 & 29.5 & 611.0 & 30 & 600.9 &  10.1
&  1.4\\
32 & 635.8 & 32.5 &  628.6 &  7.2 & 31.5 & 645.0 & 32 & 635.1 &  9.9
&  0.7\\
34 & 668.6 & 34.5 &  661.6 &  7.0 &      &       & 34 & 668.0 &
&  0.6\\
36 & 700.6 & 36.5 &  694.5 &  6.1 &      &       & 36 & 700.4 &
&  0.2\\
38 & 732.1 & 38.5 &  726.3 &  5.8 &      &       & 38 & 732.2 &  & -0.1\\
40 & 762.8 & 40.5 &  756.6 &  6.2 &      &       & 40 & 762.7 &  &  0.1\\
42 & 793.4 &      &        &      &      &       &     &       &  & \\
\end{tabular}
\end{table}

\end{document}